\documentclass{aa}
\usepackage{natbib}
\usepackage{graphicx}
\usepackage{booktabs}
\usepackage{xspace}
\usepackage{amsmath}
\usepackage{verbatim}
\usepackage{txfonts}

\bibpunct{(}{)}{;}{a}{}{,}

\newcommand{\apx}[1]{^{\rm #1}}
\newcommand{\pdx}[1]{_{\rm #1}}

\def\pdot {\dot P}
\def\pdotdot {\ddot P}
\def\edot {\dot E}

\def\nh{$N_{\rm H}$\xspace}
\def\col{cm$^{-2}$\xspace}

\def\lum{erg~s$^{-1}$\xspace}
\def\msun{~M_{\odot}}

\def\deg{^\circ}

\newcommand{\coord}[8]{R.A.\,=\,${#1}^{\rm h} {#2}^{\rm m} {#3}^{\rm s}{#4}$, Dec.\,=\,${#5}^\circ {#6}' {#7}''{#8}$\xspace}

\def\psrj{PSR J2021$+$4026\xspace}
\def\snrg{SNR G78.2$+$2.1\xspace}

\def\xmm{{\em XMM-Newton}\xspace}

\def\cha{{\em Chandra}\xspace}

\def\fermi{{\em Fermi}\xspace}
\usepackage{color}

\begin{document}

\title{Strongly pulsed thermal X-rays from a single extended hot spot on \psrj}

\author{Michela Rigoselli\inst{1}, Sandro Mereghetti\inst{1}, Roberto Taverna\inst{2,3}, Roberto Turolla\inst{3,4}, Davide De Grandis\inst{3}}

\institute {INAF, Istituto di Astrofisica Spaziale e Fisica Cosmica Milano, via A.\ Corti 12, I-20133 Milano, Italy
\and
Dipartimento di Matematica e Fisica, Universit\`a di Roma Tre, via della Vasca Navale 84, I-00146 Roma, Italy
\and
Dipartimento di Fisica e Astronomia, Universit\`a di Padova, via F. Marzolo 8, I-35131 Padova, Italy
\and
MSSL-UCL, Holmbury St. Mary, Dorking, Surrey RH5 6NT, UK
}

\offprints{michela.rigoselli@inaf.it}

\date{Received 27 October 2020 / Accepted 08 December 2020}

\authorrunning{Rigoselli et al.}

\titlerunning{Strongly pulsed thermal X-rays from \psrj}

\abstract{The  radio-quiet pulsar \psrj  is mostly known because it is the  only rotation-powered pulsar that shows variability in its $\gamma$-ray emission. Using \xmm archival data, we first confirmed that its flux is steady in the X-ray band, and then we showed that both the spectral and timing X-ray properties, i.e. the narrow pulse profile, the high pulsed fraction of 80--90\% and its dependence on the energy, can be better reproduced using a magnetized atmosphere model instead of a simply blackbody. With a maximum likelihood analysis in the energy-phase space, we inferred that the pulsar has, in correspondence of one magnetic pole, a hot spot of temperature $T\sim1$ MK and colatitude  extension $\theta\sim20\deg$. For the pulsar distance of 1.5 kpc, this corresponds to a cap of $R\sim5-6$ km, greater than the standard dimension of the dipolar polar caps.
The large pulsed fraction further argues against emission from the entire star surface, as it would be expected in the case of secular cooling. An unpulsed ($\lesssim$40\% pulsed fraction), non-thermal component, probably originating in a wind nebula, is also detected. The pulsar geometry derived with our spectral fits in the X-ray is relatively well constrained ($\chi=90\deg$ and $\xi=20\deg$--$25\deg$) and consistent with that deduced from $\gamma$-ray observations, provided that only one of the two hemispheres is active.
The evidence for an extended hot spot in \psrj, found also in  other pulsars of similar age  but not in older objects, suggests a possible age dependence of the emitting size of thermal X-rays.

}

\maketitle

\section{Introduction}
Many isolated neutron stars with characteristic ages $\tau=P/2\pdot$ in the $10^4$--$10^7$ year range show thermal X-ray emission that cannot result from their internal cooling. 
This is deduced from the small size of their emission regions and/or from their temperatures inconsistent with those predicted by neutron star cooling theories (see \citealt{pot20}, e.g., and references therein). 
Such thermal emission is instead attributed to external heating of small regions of the star surface, typically the polar caps. In fact, a fraction of the charges that flow in the magnetosphere are accelerated backward toward the star surface, heating it in localized regions and producing observable ``hot spots'' \citep{aro79,har01,har02a,che86a,che86b,chi94}. 
The study of this emission is often complicated by the presence of other spectral components: non-thermal X-rays from the magnetosphere or from spatially unresolved nebulae, and thermal emission caused by cooling of the rest of the surface, as in middle aged pulsars ($\tau\sim10^4-10^5$ years) like, for example,  the ``three Musketeers'' \citep{del05}.
When a pulsar is sufficiently old ($\tau$ $\gtrsim$ $10^6$ years) and the surface has cooled down enough, the externally-heated polar cap emission can be the only observable thermal component, as in PSR J0108$-$1431 \citep{aru19}, PSR B0943$+$10 \citep{rig19a}, PSR B1929$+$10 \citep{mis08}.
The X-ray emission from hot spots can appear significantly pulsed, if the pulsar geometry and our line of sight are favourable. Information on the angles $\chi$ and $\xi$ that the rotation axis makes with our line of sight (LOS) and with the magnetic axis, respectively, have traditionally been derived from the properties of the radio emission. More recently, constraints on the pulsar geometry have also been obtained by modeling the light curves at $\gamma$-ray \citep{2010ApJ...714..810R} and, in a few cases, X-ray energies \citep[e.g.]{2019ApJ...887L..21R}. Independent estimates of the geometrical angles will come from X-ray polarimetry (see e.g. \citealt{2014MNRAS.438.1686T,2020MNRAS.492.5057T}) thanks to the forthcoming missions \textit{IXPE} \citep{2013SPIE.8859E..08W} and \textit{eXTP} \citep{2019SCPMA..6229502Z}. These are in fact the only possibilities when radio emission is not seen, as in the radio-quiet rotation-powered pulsars (about 5\% of those  with $10^4<\tau<10^7$ years) and in other classes of objects like the X-ray-dim isolated neutron stars (XDINSs, \citealt{kap08,tur09}) and the central compact objects (CCOs, \citealt{del17}).

\setlength{\tabcolsep}{1.66em}
\begin{table*}[htbp!]
\centering \caption{Observed and derived parameters for \psrj}
\label{tab:prop}

\begin{tabular}{lccc}
\toprule
\midrule
R.A. (J2000.0)~$\apx{a}$ &		& $20^\mathrm{h}21^\mathrm{m}29^\mathrm{s}.99(3)$ & \\
Dec. (J2000.0)~$\apx{a}$ &		& $+40\deg 26' 45''.1(7)$ &  \\
$\edot$ (erg s$^{-1}$) &  & $1.2\times 10^{35}$ & \\
$B\pdx{s}$ (G)  &		& $4.0\times 10^{12}$ & \\
$\tau_c$ (yr)	&		& $73,000-77,000$ & \\
$\tau\pdx{SNR}$ (yr)~$\apx{b}$  &		& $6,600$ & \\
$d\pdx{SNR}$ (kpc)~$\apx{c}$ 	  &		& $1.5\pm0.5$ & \\

\midrule
 & LOW $\gamma$-RAY STATE & POST-RELAXATION STATE & NEW LOW $\gamma$-RAY STATE\\
& 2011 Oct -- 2014 Dec   & 2014 Dec -- 2018 Feb  & 2018 Feb -- $\dots$\\
\midrule
MJD range~$\apx{d}$ 			& $55,857-56,943$			& $57,062-57,565$ & $58,244-58,722$  \\
Epoch zero (MJD)		& $56,400$					& $57,200$ & $58,400$  \\
$P$ (s)					& $0.26532469511(1)$		& $0.26532861459(4)$  & $0.26533428586(2)$  \\
$\pdot$ (s s$^{-1}$)	& $5.7710(1)\times 10^{-14}$& $5.447(4)\times 10^{-14}$   & $5.6480(6)\times 10^{-14}$ \\
$\pdotdot$ (s s$^{-2}$)	& $+1.4(5)\times 10^{-24}$	& $-2.5(4)\times 10^{-23}$ & $+2.2(1)\times 10^{-24}$  \\
\bottomrule\\[-5pt]
\end{tabular}

\raggedright
\textbf{Notes.} Numbers in parentheses show the $1\sigma$ uncertainty for the last digits.
$\apx{a}$ The position is derived from  \citet{2011ApJS..194...17R}.
$\apx{b}$ \snrg adiabatic age is derived from \citet{2002ApJ...571..866U}.
$\apx{c}$ \snrg distance is derived from \citet{1980A&AS...39..133L}. 
$\apx{d}$ The ephemerides are derived from  \citet{2017ApJ...842...53Z,2020ApJ...890...16T}.
\end{table*}

Among radio-quiet pulsars, \psrj is of particular interest because it is the only one that exhibited flux variations at $\gamma$-rays energies.
In October 2011, \fermi-LAT observed a sudden flux drop at $E>100$ MeV, occurring over a timescale of less than one week \citep{2013ApJ...777L...2A}. This was accompanied by a significant increase in the spin-down rate (see the timing parameters in Table \ref{tab:prop}). The frequency derivative discontinuity resembles a glitch; however, the behaviour seen in \psrj differs from that of  normal glitches \citep{2011MNRAS.414.1679E,2012ApJ...755L..20P} because these are usually not associated with a flux change. Furthermore, glitches are typically followed by a rapid recovery of the timing parameters, while this was not detected for \psrj until 2014 December \citep{2017ApJ...842...53Z}. More recently, in 2018 February, \psrj entered again in a low $\gamma$-ray state, with a $\pdot$ behavior similar to the one that followed the 2011 event \citep{2020ApJ...890...16T}.

\psrj has been associated to the shell-like $\gamma$-Cygni  supernova remnant, \snrg \citep{green2009}. Its radio and X-ray shells have  a size of $\sim$1$\deg$ \citep{2013MNRAS.436..968L} and a shock velocity of $\sim$800 km s$^{-1}$ \citep{2002ApJ...571..866U}. These values, together with the SNR distance of $1.5\pm0.5$ kpc \citep{1980A&AS...39..133L}, imply an adiabatic age of 6.6 kyr, which is in agreement with the age deduced from the optical observations \citep{2003A&A...408..237M}. Thus, the age of \snrg is about one order of magnitude smaller than the spin-down age of \psrj ($\tau_c$ $\sim75$ kyr).  However, this discrepancy is not uncommon in middle-aged neutron stars (see e.g. the pulsars PSR J0538$+$2817, \citealt{2007ApJ...654..487N}, and
PSR J0855$-$4644, \citealt{2015ApJ...798...82A}, the XDINS RX J1856.5$-$3754, \citealt{2013MNRAS.429.3517M}, and the ``low-B'' magnetar SGR 0418$+$5729, \citealt{2011ApJ...740..105T}). The mismatch between the true and characteristic age can be explained if the star magnetic field substantially decayed or if the spin period at birth was close to the present one.

In the X-ray band, \psrj was frequently observed by \cha and \xmm.
Its X-ray spectrum shows a mixture of non-thermal and thermal emission: the non-thermal component, apparently non pulsed, is probably due to the pulsar wind nebula (PWN) spatially resolved by \cha \citep{2015ApJ...799...76H}; the thermal component is instead strongly pulsed  (90--100\%) with a nearly sinusoidal profile. 
\citet{2018ApJ...856...98W} analyzed two long \xmm observations, one obtained after the 2011 drop in the $\gamma$-ray flux and one in the post-relaxation state, in 2015. They could not find any significant change in the X-ray flux, spectrum or pulse profile, but noticed that the sensitivity of the current data  is not sufficient to detect the small flux change ($\sim$4\%)  expected from the observed $\pdot$ variation.

The thermal component in the spectrum of \psrj , when fitted with a blackbody model, yields an emitting region with size consistent with the dimensions of the polar cap in the dipole approximation, $R\pdx{PC}=\sqrt{2\pi R^3/Pc}\approx300$~m \citep{2015ApJ...799...76H,2018ApJ...856...98W}. 
Given the viewing angle of $\chi$ $\sim$ $90\deg$ inferred by the $\gamma$-ray data \citep{2010MNRAS.405.1339T}, the two polar caps are visible. If both contributed to the X-ray emission, then the pulse profile would be far less pulsed ($\lesssim$25\%) and, for several values of $\xi$, it would have two peaks \citep[e.g.]{bel02}. \citet{2015ApJ...799...76H} concluded that the strongly pulsed and single-peaked profile implies that only one of the two polar caps is active in X-rays.  

Here we present a reanalysis of the \xmm data aimed to quantitatively reproduce the timing and the spectral features of \psrj fitting its phase-resolved spectrum with magnetized atmosphere models.
n our spectral models,  specifically computed for this pulsar, the emitting region does not have to be point like and the cap semi-opening angle $\theta\pdx{cap}$ can take any value from $0\deg$ to $90\deg$. The temperature and the magnetic field at each latitude are consistently evaluated considering a dipolar magnetic field.

The paper is organized as follows. In section~\ref{sec:model} we briefly describe our computation of the magnetized atmosphere models, and we illustrate the approach we used to compute the phase-dependent spectrum emitted by the pulsar. We then describe the data analysis (section~\ref{sec:data}) and apply our model to the observed phase-averaged and phase-resolved spectra (section~\ref{sec:spectra}), and to the pulse profiles (section~\ref{sec:timing}). The results are discussed in section~\ref{sec:disc}.

\section{Modeling the X-ray pulse profiles and spectra}
\label{sec:model}

Our computation of the phase-dependent spectrum emitted by a neutron star, as seen by a distant observer, is done in four steps: i) defining the stellar parameters (mass and radius, temperature and magnetic field, geometry of the pulsar); ii) evaluating the local spectrum emitted by each patch of the surface; iii) collecting  the contributions of all surface elements that are in view at different rotation phases, accounting for general relativistic effects, such as redshift and light bending; iv) convolving the observed flux at infinity with the instrumental response matrix, in order to perform spectral and timing analysis.

We adopted realistic values of mass $M=1.36\msun$ and radius $R=13$ km, that are consistent with the most recent equation of states (EOSs, see e.g. \citealt{2016PhR...621..127L} and references therein) and give a gravitational redshift factor $z\sim0.2$.
We computed the model atmosphere at 10 different values of the co-latitude (equally spaced in $\mu=cos\theta$ so that they have the same area), assuming a dipole magnetic field and the ensuing temperature distribution \citep{gre83}. We considered models with two values of magnetic field and temperature at the poles: [$T\pdx{p}=1$ MK, $B\pdx{p}=4\times 10^{12}$ G] and  [$T\pdx{p}=0.5$ MK, $B\pdx{p}=3\times 10^{12}$ G].
The geometrical angles, $\chi$ and $\xi$, that the LOS and the dipole axis make with rotation axis, respectively, are sampled by means of $19\times 19$ equally-spaced grid ranging from $0\deg$ to $90\deg$ each. 

The atmospheric structure and radiative transfer have been computed using the code developed by \citet[see also \citealt{2003ApJ...593.1024L,2006MNRAS.366..727Z,gon16}]{2003astro.ph..3561L}, which applies the complete linearization technique to the case of a semi-infinite, plane-parallel atmosphere in radiative equilibrium. Radiation transfer calculations are performed accounting for strong magnetic fields,  solving the radiative transfer equation for photons polarized both in the ordinary (O) and in the extraordinary (X) modes, with electric field oscillating either parallel or perpendicular, respectively, to the plane made by the photon propagation direction and the local magnetic field \citep{gin70,mes92}. 
Opacities are evaluated accounting for magnetic effects. Although the code can be generalized to mixed Hydrogen-Helium compositions and extended to the case of partial ionization, for the sake of simplicity we restrict here only to a pure-Hydrogen, fully ionized atmospheric slab. Each run requires in input the intensity $B$ of the local magnetic field and the angle $\theta_{\rm B}$ it forms with the local slab normal, the effective temperature $T$ and the surface gravity $g$. For this reason, we divided the atmospheric layer into a number of plane-parallel patches, infinitely extended in the transverse direction and emitting a total flux $\sigma T^4$. The code returns in output the intensities $I\pdx{O}$ and $I\pdx{X}$ of the emerging O- and X-mode photons as functions of the energy $E$ and the two polar angles $\theta_k$ and $\phi_k$ which identify the photon direction $\mathbf{k}$ with respect to the local normal. These angles can be written as functions of the two viewing angles $\chi$ and $\xi$ and of the so-called ``impact angle'' $\eta$, which provides the inclination of the magnetic axis wrt the LOS at each rotational phase $\gamma=\Omega t$:
\begin{equation}
   \cos\eta = \cos\xi \cos\chi + \sin\xi \sin\chi \cos\gamma  
    \label{eq:eta}
\end{equation}
(see e.g. \citealt{tav15,gon16}).

The radiation transfer equation was solved for 20 photon energies uniformly distributed on a logarithmic scale from 0.1 to 10 keV, for 10  values of $\alpha=\cos\theta\pdx{B}$, 15 values of  $\mu_k=\cos\theta_k$, and 5 values of $\phi_k$, all linearly spaced. The southern hemisphere is built exploiting the symmetry properties of the opacities: $\mu_k = -\mu_k$, $\phi_k = \pi-\phi_k$.

Once the emerging flux at each patch is known, the spectrum at infinity is computed by collecting all the contributions that are in view at a certain rotational phase $\gamma$, accounting for general relativistic effects. Since we are interested in radiation emerging from polar caps which are not necessarily point-like, we considered semi-opening angles  $\theta\pdx{cap}$ of $5\deg$ (which can still be treated as point-like), $10$, $20$, $30$, $45$, $65$, and $90\deg$ (the whole hemisphere).
The observed flux was stored in a seven-dimensional array $F^\infty\,(E,$ $\gamma,$ $B\pdx{p},$ $T\pdx{p},$ $\theta\pdx{cap},$ $\xi,$ $\chi)$, which associates at each set of parameters the (discrete) values of the energy- and phase-dependent intensity.

The final step of the computation consisted in convolving the array $F^\infty$ with the instrumental response, in order to properly compare the model with the observed data, as described in the next section.

\section{Observations and data analysis} \label{sec:data}

\setlength{\tabcolsep}{0.5em}
\begin{table*}[htb!]
\centering \caption{Exposure times and source counts for \psrj in the three EPIC cameras.}
\label{tab:counts}

\footnotesize
\begin{tabular}{lccccccc}
\toprule
\midrule
Obs. ID & Start time & End time  & Camera    &  Net Exposure    & Operative Modes$\apx{a}$  & Source Counts$\apx{b}$  & Count Rate$\apx{b}$   \\[3pt]
       &  UT & UT &    & ks          &  and filters  &                    cts                     & $10^{-3}$ cts/s  \\[3pt]
\midrule\\[-5pt]
0670590101  & 2012-04-11 07:24:03 & 2012-04-12 21:07:38    & pn    & 63.7  & SW, medium    & $778 \pm 41$ & $12.2 \pm 0.6$ \\[3pt]
 & & & MOS1  & 88.1   & FW, medium    & $374 \pm 28$ & $4.2 \pm 0.3$ \\[3pt]
 & & & MOS2  & 94.7   & FW, medium    & $402 \pm 30$ & $4.2 \pm 0.3$ \\[3pt]

\midrule\\[-5pt]
0763850101  & 2015-12-20 10:15:58 &	2015-12-22 01:24:17    & pn    & 90.0  & SW, medium   & $1127 \pm 50$ & $12.5 \pm 0.6$ \\[3pt]
 & & & MOS1  & 126.2  & FW, thick     & $305 \pm 26$ & $2.4 \pm 0.2$ \\[3pt]
 & & & MOS2  & 125.9  & FW, thick     & $397 \pm 29$ & $3.4 \pm 0.3$ \\[3pt]

\midrule\\[-5pt]
sum      & &  & pn    & 153.7  & & $1905 \pm 65$ & $12.4 \pm 0.4$ \\[3pt]
& & & MOS1  & 214.4  & &  $679 \pm 39$ &  $3.2 \pm 0.2$ \\[3pt]
& & & MOS2  & 220.6  & &  $830 \pm 43$ &  $3.8 \pm 0.2$ \\[3pt]
\bottomrule\\[-5pt]
\end{tabular}

\raggedright
\textbf{Notes.} $\apx{a}$ SW = Small Window; FW = Full Window. $\apx{b}$ Net counts and count rate extracted with the ML in the energy range $0.7-3$ keV.
\end{table*}

We analyzed the two longest \xmm observations of \psrj, that were obtained in 2012 April (Obs. ID 0670590101) and in 2015 December (Obs. ID 0763850101), in the low $\gamma$-ray and in the post-relaxation states, respectively (see Table \ref{tab:counts}).
The MOS1/2 cameras were operated in full-window mode with medium and thick optical filter, while the pn camera was in small-window mode with medium filter. Only the pn time resolution (5.7 ms wrt 2.6 s of the MOS cameras) is adequate to reveal the pulsations of the source.

The data reduction was performed using the \textsc{epproc} and \textsc{emproc} pipelines of version 15 of the Science Analysis System (SAS). We selected single- and multiple-pixel events (\textsc{pattern}$\leq$4 and $\leq$12) for both the pn and MOS1/2. We then removed time intervals of high background using the SAS program \textsc{espfilt} with standard parameters. The resulting net exposure times and counts are summarized in Table~\ref{tab:counts}.

For all the timing analysis, we folded the data at the periods derived from the known pulsar ephemeris appropriate for each observing epoch (Table~\ref{tab:prop}), after correcting the time of arrivals to the Solar System barycenter with the tool \textsc{barycen}.

\subsection{Maximum Likelihood spectral extraction}

To extract the source counts and spectra, we used a maximum likelihood (ML) technique, as implemented by \citet{rig18,rig19a}. In short, this consists in estimating the most probable number of source and background counts that reproduce the observed data, assuming that source events are spatially distributed according to the instrumental point-spread function (PSF), while the background events are uniformly distributed. The expectation value of total counts in the image pixel $(i,j)$ is 
\begin{equation}
\mu_{ij} = b + s \times \mathrm{PSF}_{ij},
\label{eq:mu2D}
\end{equation}
where $b$ gives the background in counts per unit area (cts asec$^{-2}$),  $s$ is the total number of source counts, and  $\mathrm{PSF}_{ij}$ is the normalized point-spread function corresponding to that pixel.
We take into account the PSF dependence on photon energy and position on the detector, as  derived from in-flight calibrations \citep{simo}. The ML method has the advantage to exploit all the source events that are located in the region of interest and compatible with the PSF. Furthermore, the background is determined locally, and not in a different region of the detector.

The above ML method, that from this point on will be dubbed as ``2D-ML'', can be generalized to take into account also  the pulse phase information of the events for periodic sources (``3D-ML'', \citealt{her13}). If the events are binned in spatial and phase coordinates, a tridimensional space is defined, where the expectation value of the bin $(i,j,k)$ is 
\begin{equation}
\mu_{ijk} = b + s\pdx{u} \times \mathrm{PSF}_{ij} + s\pdx{p} \times \mathrm{PSF}_{ij} \times f_k.
\label{eq:mu3D}
\end{equation}
Now $s\pdx{u}$ and $s\pdx{p}$ represent the source counts for the unpulsed and pulsed components, respectively, while $f_k$ is the normalized pulse profile at phase $\varphi_k$. In this work, to describe the pulse shape we considered a sine function
\begin{equation}
f_k=1+\sin(\varphi_k-\varphi_0),
\label{eq:sin}
\end{equation}

\noindent
and a Gaussian function
\begin{equation}
f_k=\frac{1}{\sqrt{2\pi\sigma_\varphi^2}}\exp{\left[-\frac{(\varphi_k-\varphi_0)^2}{2\sigma_\varphi^2} \right]},
\label{eq:gauss}
\end{equation}
where $\varphi_0$ is the absolute phase and $\sigma_\varphi$ is the characteristic Gaussian width.

The maximum likelihood ratio (MLR), defined as the difference between the likelihood of the assumed model with its best parameters and that of the null hypothesis, is used to evaluate the significance of the results. In the case of 2D-ML, the MLR compares the likelihood of having a point source with respect to having only background, while in the case of 3D-ML, it is obtained as a comparison between the likelihood of having a pulsating source with respect to  the likelihood of having a source with constant emission.
The significance in $\sigma$ of the detection is the square root of the MLR.

Source spectra can be extracted by applying the ML to the images in different energy bins (using for each one the appropriate PSF). 
If we use eq. \ref{eq:mu2D} with the whole dataset, we get the phase-averaged spectrum, while if we divide the data into phase bins, we get the phase-resolved spectra. Note that all the spectra derived in this way contain the contributions of both the pulsed and unpulsed emission. Conversely, using the 3D-ML (eq. \ref{eq:mu3D}) we can obtain distinct  spectra for the pulsed and unpulsed emission. 
In this context, the pulsed fraction (PF) is defined as the ratio between the pulsed and the total counts, as a function of the energy:
\begin{equation}
    \mathrm{PF}(E) = \frac{s\pdx{p}(E)}{s\pdx{u}(E)+s\pdx{p}(E)}.
    \label{eq:PF}
\end{equation}

\begin{figure*}[htb!]
  \centering
  \includegraphics[width=1.\linewidth]{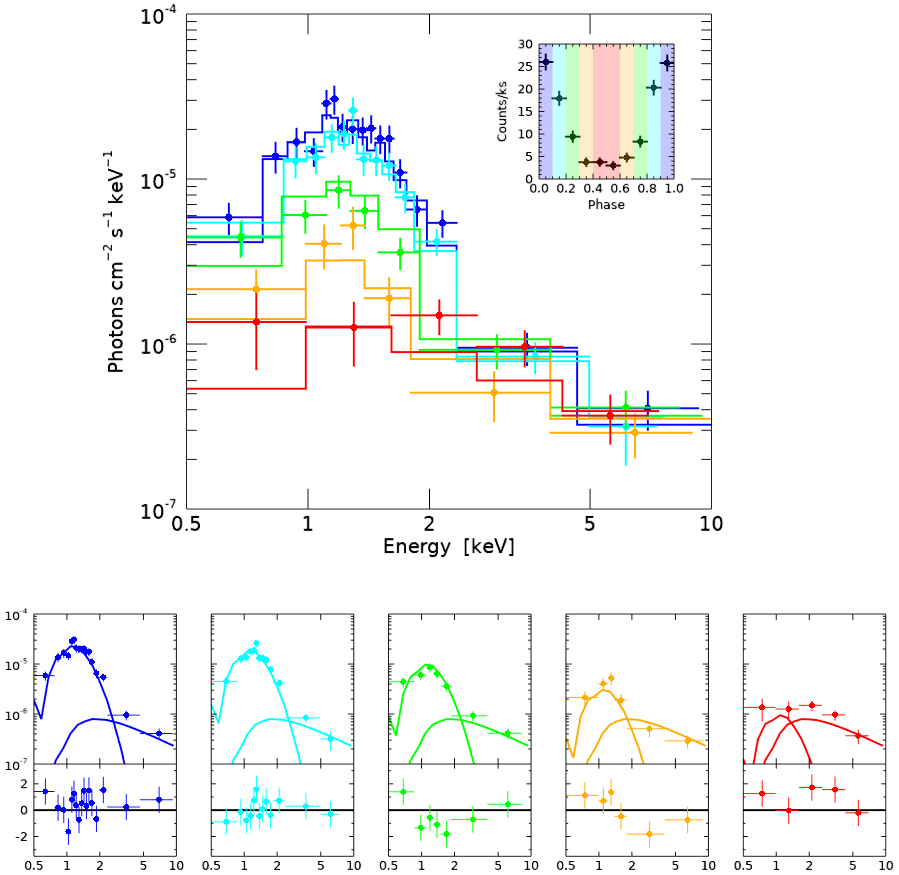}
  \caption{Phase-resolved spectrum of \psrj fitted with a power law and the atmosphere model discussed in section \ref{sec:model} ($T\pdx{p}=1$ MK, $B\pdx{p}=4\times10^{12}$ G, $\theta\pdx{cap}=20\deg$, $\chi=90\deg$, $\xi=25\deg$). The main panel shows the  spectra corresponding to the five phase bins, displayed in the inset.
  The bottom panels show the five spectra and their residuals wrt the best fit (same color code as before). The lines indicate the two spectral components:  the thermal one  changes with phase, while the power law is constant.}
  \label{fig:spectra2d}
\end{figure*}

We applied the ML analysis in a circular region centered at \coord{20}{21}{32.}{5}{+40}{26}{46.}{04} with a radius of $40''$, in order to account for the different fields of view of each observation and camera. 
We first extracted the pulse profile dividing the data set into ten phase bins. Then, considering that the folded light curves are symmetric around phase 0.5 and also the atmosphere model has the same symmetry, we summed the corresponding spectra two by two to obtain the five phase ranges highlighted in the inset of Figure~\ref{fig:spectra2d}.

\subsection{Spectral fitting}

We performed the spectral fits using XSPEC (version 12.11.0)  and the photoelectric absorption model \textsc{tbabs}, with cross sections and abundances from \citet{wil00}. The spectra of the three cameras were fitted simultaneously, including a normalization factor to account for possible cross-calibration uncertainties.
All the spectra were grouped to achieve at least 100 (phase-averaged spectra) or 50 counts (phase-resolved and unpulsed/pulsed spectra) in each energy bin. 
We give all the errors at $1\sigma$ confidence level. 

Phase-resolved spectroscopy is usually performed by fitting independently the spectra corresponding to different phase intervals and examining how  the derived best-fit parameters change as a function of the phase. In our analysis we followed a different approach, based on global fits in the   energy-phase space. In fact, by proper integration of the array $F^\infty$ described in section \ref{sec:model} over $E$ and $\gamma$, we can obtain the model flux in each energy and phase bin. This model is fitted simultaneously to the phase-resolved spectra to derive a single set  of the best-fit parameters $B\pdx{p},$ $T\pdx{p},$ $\theta\pdx{cap},$ $\xi,$ $\chi$.

\section{Results} \label{sec:res}

We first applied the ML to the two single  observations and we found no significant variations between the two epochs: the count rate measured by the pn camera is $0.0122\pm0.0006$ cts s$^{-1}$ in 2012, and $0.0125\pm0.0006$ cts s$^{-1}$ in 2015, a difference of $2.4\pm6.8 \%$. 
We can thus set a $3\sigma$ upper limit of $22.8 \%$ on the long-term variability.
The count rate measured by the MOS cameras are different in the two observations (see Table~\ref{tab:counts}), but this is due to the different setting of the instruments in the two epochs.
Therefore, in the following, we will present the results obtained by summing the data the two observations\footnote{We summed independently the spectra of the three cameras; the folded light curves were added after an appropriate phase shift to align the pulse profiles.}.

\subsection{2D-ML spectral analysis} \label{sec:spectra}

We first modeled the phase-averaged spectrum of \psrj using a power law plus a blackbody, and we found that this model fits well the data, with a reduced $\chi_\nu^2=0.80$ for 32 degrees of freedom (dof), null-hypothesis probability (nhp) of 0.78. The best-fitting photon index $\Gamma=1.2\pm0.2$, observed temperature $kT^\infty=0.221\pm0.015$ keV, and emitting radius $R\pdx{em}^\infty=340_{-80}^{+110}$ m (evaluated for $d=1.5$ kpc) are in excellent agreement with what found in previous works \citep{2015ApJ...799...76H,2018ApJ...856...98W}. For this model, the column absorption is \nh$=7.1_{-0.8}^{+0.9} \times 10^{21}$ \col.

We then fitted the spectrum with some of the magnetized hydrogen atmosphere models available in XSPEC. In particular, the \textsc{nsmaxg} models \citep{ho08,ho14} allow to specify if the magnetic field and the temperature are constant on the surface, or they follow the profile expected for a dipole field. 
As summarized in the first part of Table~\ref{tab:spec}, all these models give a good fit if they are combined with an absorbed power law with $\Gamma \sim 1.1$ and  \nh$\sim 9 \times 10^{21}$ \col.
The effective temperature is about 0.66 MK or 1 MK, depending on whether the impact angle $\eta$ (see eq. \ref{eq:eta}) is $0\deg$ or $90\deg$, respectively. The corresponding emitting radii $R\pdx{em}$ of $15.6_{-6.9}^{+10.0}$ km or $5.9_{-2.6}^{+3.8}$ km, indicate that the thermal emission comes from a very large region or even from the whole surface; in fact, if we fix $R\pdx{em}=R=13$ km, we still obtain a good fit.

To further investigate the  possible emission from the whole surface, we added to the best-fitting spectrum a second thermal component with temperature $T\pdx{cool}$ and  emitting radius fixed  to that of the star. We let $T\pdx{cool}$ free to vary, but the resulting best-fitting value was not constrained. In the case of blackbody models, we found $T\pdx{cool}<84$ eV with $\chi^2_\nu=0.69$ for 31 dof (corresponding to an F-test probability of 0.02 compared to the single-blackbody model). We repeated the analysis with the \textsc{nsmaxg} models and found $T\pdx{cool}<63$ eV, with $\chi^2_\nu=0.89$ for 31 dof  (F-test probability of 0.85). 
These results indicate that the addition of a thermal component from the whole surface is not statistically required and we derive  a $3\sigma$ upper limit of its luminosity in the range $(5-16)\times10^{32}$ \lum (depending on the thermal model used).

Finally, we used our magnetized atmosphere models presented in section~\ref{sec:model} and we found that only the model with $T\pdx{eff}=1$ MK gives an acceptable $\chi^2_\nu$ (0.96 for 33 dof wrt 3.8 of the model with $T\pdx{eff}=0.5$ MK). All the explored $\chi$ and $\xi$ angles gave  equally good results, for best fit emitting regions with $\theta\pdx{cap}\sim20\deg$, that   corresponds to a radius of about 5$-$6 km. The addition of an absorbed power law with $\Gamma=1.0\pm0.2$ and \nh$=(8.5\pm0.4) \times 10^{21}$ \col is required also in these cases.

To constrain the pulsar geometry, we had to rely on the phase-resolved spectroscopy: we fitted the five phase-resolved spectra of \psrj with our models for all the geometries and we got good fits ($\chi^2_\nu<2$) for a restricted set of angles: $100\deg \lesssim \chi+\xi \lesssim 120\deg$. Moreover, as previous works have already noticed, it is impossible to reproduce the observed data if X-rays are emitted by both hemispheres. The best fit is obtained when $\chi=90\deg$ and $\xi=25\deg$, with approximately the same spectral parameters found in the phase-averaged spectral analysis, see Table~\ref{tab:spec}. We remark that with our fitting method, the normalization of the atmosphere models is linked for all the phases and it leads to $R\pdx{em}=5.1\pm0.2$ km or $R\pdx{em}=5.5\pm0.2$ km, depending on which of the two  hemispheres is active. 
These results were obtained including a power law with constant flux at all phases, consistent with the assumption that the non-thermal component is entirely unpulsed.

\subsection{3D-ML timing and spectral analysis} \label{sec:timing}

\begin{figure*}[htb!]
  \centering
  \includegraphics[width=0.49\linewidth]{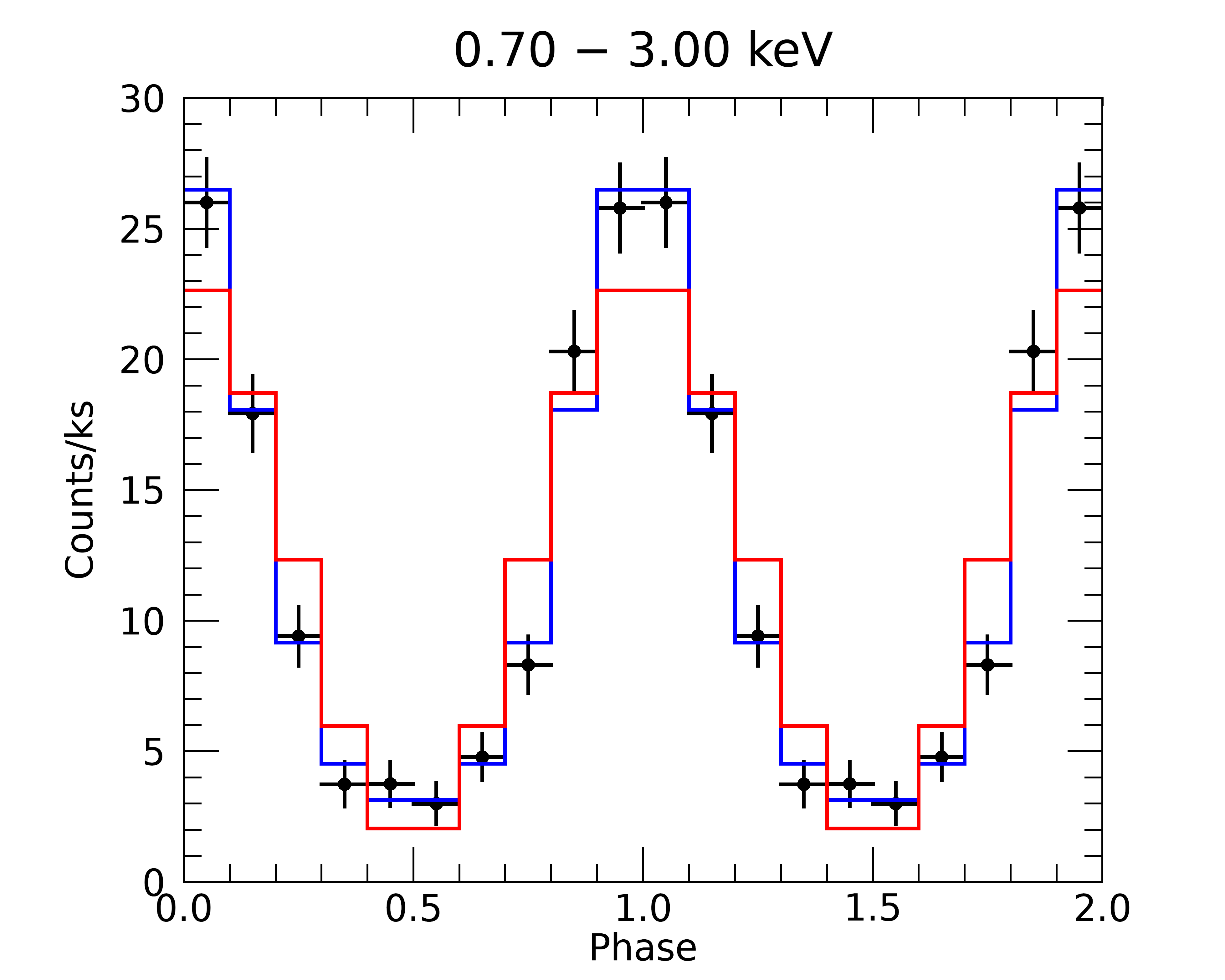}
  \includegraphics[width=0.49\linewidth]{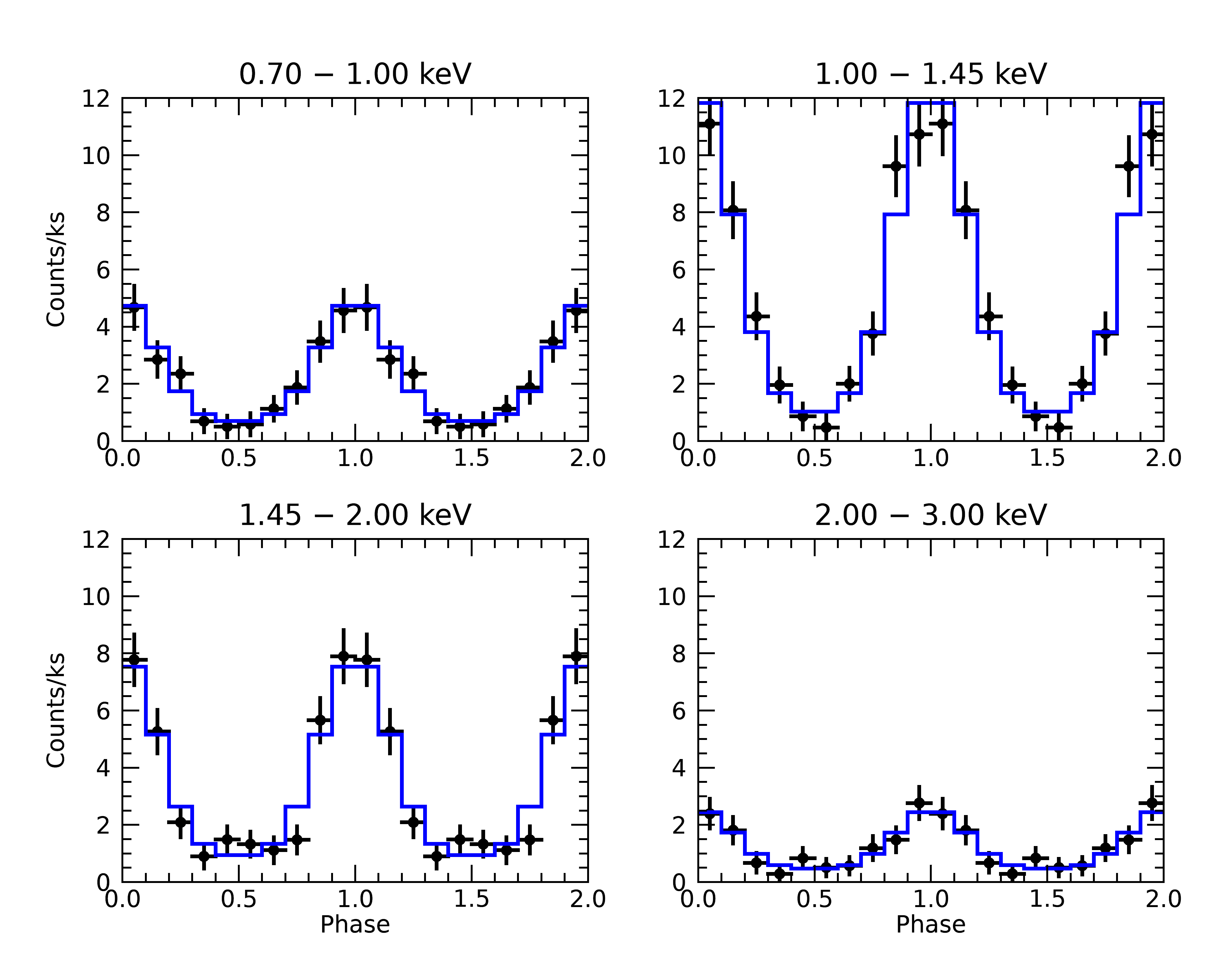}
  \caption{Phase-folded light curves of \psrj as observed by EPIC-pn in the range of $0.7-3$ keV (left panel) and $0.7-1-1.45-2-3$ keV (right panel). The data (black dots) are obtained with the 2D-ML in ten phase bins. The red (sine, eq. \ref{eq:sin}) and blue (Gaussian, eq. \ref{eq:gauss}) lines are obtained with the 3D-ML.}
  \label{fig:pulse}
\end{figure*}

The 3D-ML analysis allows us to simultaneously exploit in a very effective way the combined timing and spectral information. 
The pulsations of \psrj are detected with the highest significance in the  $0.7-3$ keV energy range. 
Applying the 3D-ML analysis in this range with 10 phase bins, we found that the pulse profile is described better by a Gaussian (MLR$\pdx{gauss}=245$) than by a sine function (MLR$\pdx{sine}=227$), as it is shown in the left panel of Figure \ref{fig:pulse}. The use of a Gaussian thus gives an improvement of the MLR of 18, corresponding to a significance greater than $4\sigma$.

\begin{figure*}[htb!]
  \centering
  \includegraphics[width=0.49\linewidth]{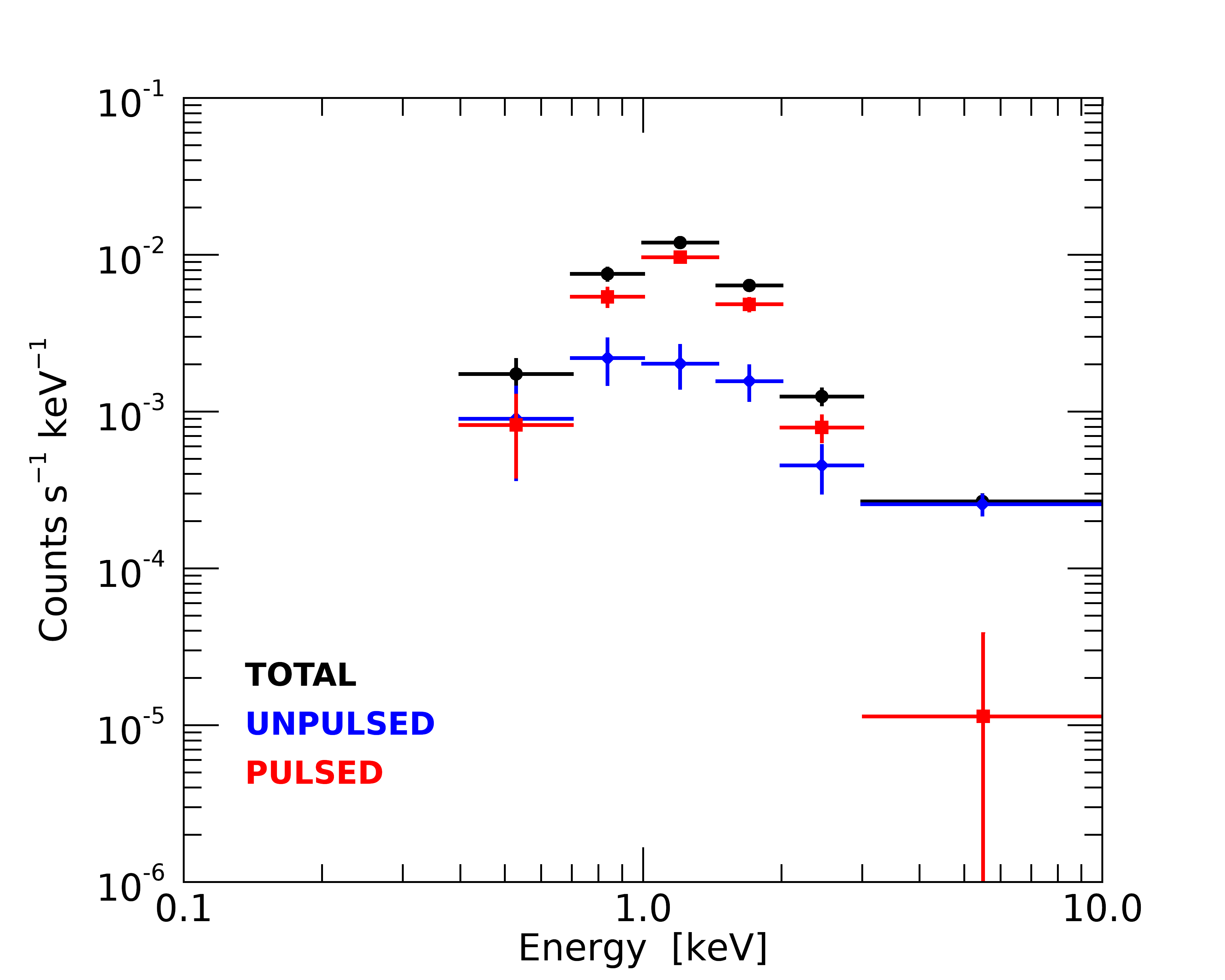}
  \includegraphics[width=0.49\linewidth]{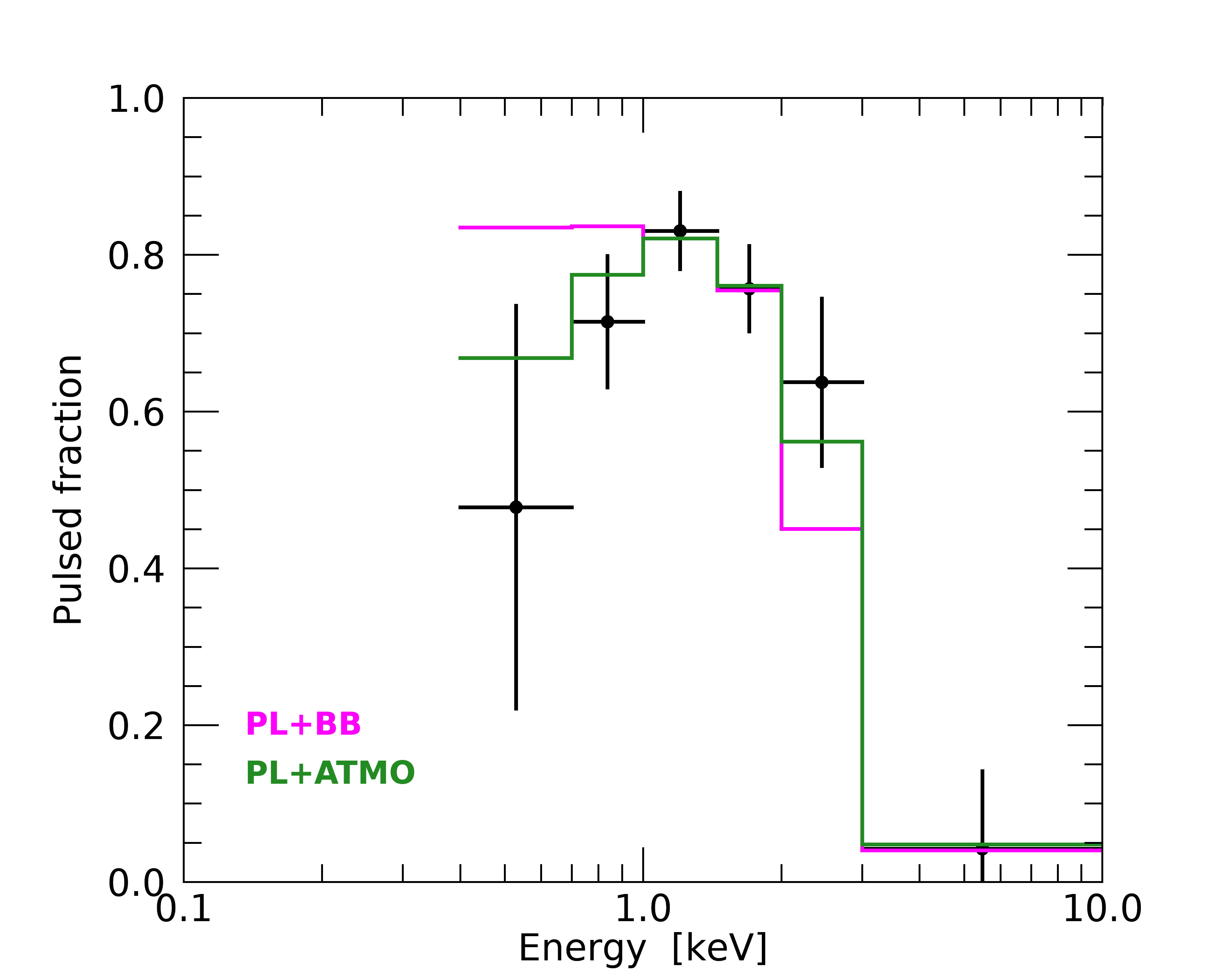}
  \caption{Total, unpulsed and pulsed spectra (left panel) and corresponding PF (right panel) of \psrj as observed by EPIC-pn and extracted with the 3D-ML, assuming for the pulse profile a Gaussian function (eq. \ref{eq:gauss}). 
  The solid lines represent the PF computed in the case of a power law plus a blackbody (PL+BB, magenta line), and a power law plus a magnetized atmoshpere (PL+ATMO, green line). }
  \label{fig:spectra3d}
\end{figure*}

The Gaussian has a phase width $\sigma_\varphi=0.15 \pm 0.01$, and PF=$0.77 \pm 0.05$ (defined as in eq.~\ref{eq:PF}).
In the softer energy range ($0.4-0.7$ keV) we found hints for pulsations, with PF=$0.48\pm0.26$, at a $1.5\sigma$ level, while above 3 keV no pulsations are detected.In fact, the source
has $s=288\pm28$ counts above 3 keV, corresponding to a detection at more than $10\sigma$, but only $s\pdx{p}=11_{-11}^{+28}$ of these counts are pulsed, yielding a $3\sigma$ upper limit PF$<$0.34.
Then, we divided the central energy range into four bins ($0.7-1-1.45-2-3$ keV) and we applied the 3D-ML analysis to each pulse profile with a Gaussian $\sigma_\varphi$ fixed at 0.15. The data and the corresponding best fits are shown in Figure \ref{fig:pulse}, right panel.
The measured $s\pdx{u}(E)$ and $s\pdx{p}(E)$ where used to derive the unpulsed and the pulsed spectra, respectively, and the PF as a function of energy (see Figure~\ref{fig:spectra3d}). 

We fitted the unpulsed and pulsed spectra with our models, with the respective normalizations correctly evaluated as explained in section \ref{sec:model}. Differently from what we did in the previous section, now we can relax the assumption that the power law has a constant flux, and we can investigate its contribution to the pulsed spectrum simply by adding a power law model with free normalization to each spectrum. We found that the best-fitting geometry is $\chi=90\deg$ and $\xi=20\deg$, and that the \textit{unpulsed} power law has a normalization consistent with 0,
independently of which of the two hemispheres is emitting. We obtained $\chi^2_\nu=1.03$ and 1.09 for 9 dof, respectively, and spectral parameters very similar to those found with phase-resolved spectroscopy. We also tested the power-law plus blackbody model, adopting the same hypothesis that both the unpulsed and the pulsed spectra could show a mixture of thermal and non-thermal emission. We found a worse best fit, with $\chi^2_\nu=1.61$ for 7 dof. Also in this case the power law is entirely unpulsed, while the blackbody contributes to both the unpulsed and the pulsed spectra.

Using the best-fit parameters of all the models summarized in the last part of Table~\ref{tab:spec}, we computed the expected PF as a function of energy, that  is shown in Figure~\ref{fig:spectra3d}, right panel. Above 3 keV, the observed emission is entirely due to the non-thermal photons.
The best-fit normalizations  of the power-law components imply a $3\sigma$ upper limit of $\sim$0.40 on the PF above 3 keV, independent of the specific thermal emission model.

\setlength{\tabcolsep}{0.3em}
\begin{table*}[htbp!]
\centering \caption{Spectral results}
\label{tab:spec}

\begin{tabular}{lcccccccccc}

\toprule
\midrule

Model				& \nh			& $\Gamma$	& PL normalization		& $B\pdx{p}\apx{~a}$	& $\chi\apx{~a}$	& $\xi\apx{~a}$	& 

$T\pdx{eff}$	& $R\pdx{em}$	& $\chi_\nu^2$/dof	& nhp\\[5pt]
            		& $10^{21}$ \col&			& $10^{-6}$ pho cm$^{-2}$ s$^{-1}$ keV$^{-1}$ & $10^{12}$ G	& $\deg$& $\deg$	&
            		 
MK		& km	&					& 	 \\[5pt]
\midrule\\
\multicolumn{10}{c}{\large PHASE AVERAGED}\\
\midrule\\[-5pt]

PL+BB	& $7.1_{-0.8}^{+0.9}$	& $1.2\pm0.2$			& $3.5_{-1.1}^{+1.5}$	& \dots & \dots & \dots &
$3.1\pm0.2$		& $0.28_{-0.07}^{+0.09}$& $0.80/32$	& 0.78 \\[5pt]

PL+NSMAXG$\apx{~b}$	& $9\pm1$	& $1.1\pm0.3$	& $2.8_{-1.0}^{+1.5}$	& $4$	& \dots & \dots  &
$1.1 \pm 0.1$		& $4.4_{-1.7}^{+3.2}$	& $0.86/32$	& 0.69 \\[5pt]

PL+NSMAXG$\apx{~c}$	& $9\pm1$	& $1.2_{-0.3}^{+0.2}$	& $3.2_{-1.1}^{+1.5}$	& $2$	& \multicolumn{2}{c}{$\eta=0\deg$} &
$0.66_{-0.09}^{+0.10}$& $15.6_{-6.9}^{+10.0}$	& $0.84/32$	& 0.72 \\[5pt]

PL+NSMAXG$\apx{~c}$	& $9\pm1$	& $1.1_{-0.3}^{+0.2}$	& $2.9_{-1.1}^{+1.4}$	& $2$	& \multicolumn{2}{c}{$\eta=90\deg$} &
$1.1_{-0.1}^{+0.2}$& $5.9_{-2.6}^{+3.8}$	& $0.86/32$	& 0.69 \\[5pt]

PL+NSMAXG$\apx{~c}$	& $10.7\pm0.4$	& $1.3\pm0.2$	& $4.2_{-1.1}^{+1.4}$	& $2$	& \multicolumn{2}{c}{$\eta=90\deg$} &
$0.88\pm0.01$& $13\apx{~a}$	& $0.76/33$	& 0.64 \\[5pt]

PL+ATMO north $\apx{~d}$ & $8.4\pm0.4$ & $1.0\pm0.2$ & $2.3_{-0.6}^{+0.8}$ & $4$ & $90$ & $25$ & $1\apx{~a}$ & $5.1\pm0.2$ &  0.96/33  & 0.53  \\[5pt]

PL+ATMO south $\apx{~d}$ & $8.5\pm0.4$ & $1.0\pm0.2$ & $2.3_{-0.6}^{+0.8}$ & $4$ & $90$ & $25$ & $1\apx{~a}$ & $5.6\pm0.2$ &  0.96/33  & 0.53  \\[5pt]

\midrule\\
\multicolumn{10}{c}{\large PHASE RESOLVED}\\
\midrule\\[-5pt]

PL+ATMO north $\apx{~d}$ & $8.3\pm0.4$ & $1.0\pm0.3$ & $2.2_{-0.7}^{+0.9}$ & $4$ & $90$ & $25$ & $1\apx{~a}$  & $5.1\pm0.2$ &  1.05/44  & 0.38  \\[5pt]

PL+ATMO south $\apx{~d}$  & $8.3\pm0.4$ & $1.1\pm0.3$ & $2.8_{-0.9}^{+1.1}$ & $4$ & $90$ & $25$ & $1\apx{~a}$ & $5.5\pm0.2$ &  0.94/44  & 0.59  \\[5pt]

\midrule\\
\multicolumn{10}{c}{\large UNPULSED / PULSED}\\
\midrule\\[-5pt]

PL+BB & $6\pm1$ & $0.92\pm0.26$ & $1.9_{-0.7}^{+0.9}$ / $<$0.4 & $\dots$ & $\dots$ & $\dots$ & $3.6\pm0.3$ & $0.06_{-0.02}^{+0.03}/0.16_{-0.04}^{+0.06}$ &  1.61/7  & 0.13   \\[5pt]

PL+ATMO north $\apx{~d}$ & $8.7\pm0.6$ & $1.0\pm0.2$ & $2.1_{-0.6}^{+0.8}$ / $<$0.4 & $4$ & $90$ & $20$ & $1\apx{~a}$ & $5.5\pm0.3$ &  1.03/9  & 0.41 \\[5pt]

PL+ATMO south $\apx{~d}$  & $8.8\pm0.6$ & $1.1\pm0.2$ & $2.4_{-0.7}^{+0.8}$ / $<$0.5 & $4$ & $90$ & $20$ & $1\apx{~a}$ & $5.7\pm0.3$ &  1.11/9  & 0.35  \\[5pt]

\bottomrule\\[-5pt]
\end{tabular}

\raggedright
\textbf{Notes.} 
Temperatures and radii are at the star surface; $M=1.36\msun$, $R=13$ km, $d=1.5$ kpc. 
Errors at $1\sigma$.
$\apx{a}$ Fixed value.
$\apx{b}$ \textsc{nsmaxg} model \citep{ho08,ho14} with constant magnetic field and surface temperature.
$\apx{c}$ \textsc{nsmaxg} model with a dipole distribution of the magnetic field and consistent temperature distribution.
$\apx{d}$ Our model of magnetized atmosphere described in section~\ref{sec:model} with a dipole distribution of the magnetic field and consistent temperature distribution, with only either the northern or the southern hemispheres active.
\end{table*}

\section{Discussion} \label{sec:disc}

We have shown that the use of a magnetized atmosphere model instead of a blackbody provides a better explanation of the observed X-ray properties of \psrj. In particular, the energy dependence of the PF is reproduced better by our model, as it is shown in Figure~\ref{fig:spectra3d}, right panel. Another weakness of the blackbody model is that it predicts a sinusoidal pulse profile, but our analysis clearly indicates a narrower pulse, well described by a Gaussian shape with $\sigma_\varphi=0.15$ in phase (Figure~\ref{fig:pulse}, left panel). 

Both the spectral and timing properties of \psrj can be well reproduced using our hydrogen atmosphere model with $T\pdx{p}=1$~MK, $B\pdx{p}=4\times10^{12}$~G and $\theta\pdx{cap}$ $\sim$20$\deg$,  provided that, as in previous works \citep{2015ApJ...799...76H,2018ApJ...856...98W}, one of the two magnetic polar regions does not emit detectable X-rays. 
The deactivation of a polar cap is possible in outer gap models. In fact, due to the gravitational deflection of some of the high-energy photons emitted by the primary charges and to the local multipolar magnetic field, the charges can fill one of the gaps and quench the accelerator zone \citep{2000ApJ...537..964C}.
The pulsar geometry derived from our X-ray fits is relatively well constrained ($\chi=90\deg$ and $\xi=20\deg$--$25\deg$) and consistent with that deduced from $\gamma$-ray observations \citep{2010MNRAS.405.1339T}. 
The thermal emission has a bolometric luminosity of $(4.6\pm0.3)\times10^{31}$ \lum (for $d=1.5$ kpc).
Non-thermal emission with a luminosity of $L_{1-10}=(9.2\pm0.7)\times10^{30}$ \lum, corresponding to $7.7\times10^{-5}$ times the spin-down power, is also present and we set a $3\sigma$ upper limit of $\sim$40\% on its PF. This component arises from the non-thermal particles accelerated in the outer magnetosphere or, more likely, in the PWN resolved by \cha \citep{2015ApJ...799...76H}.

If \psrj is really the remnant of \snrg, its small true age of about 7 kyr implies that its surface should be still hot enough to significantly emit in the X-ray band. 
To check this possibility, we added to the best-fitting spectrum a second thermal component with temperature $T\pdx{cool}$ and fixed emitting radius equal to that of the star.
We let $T\pdx{cool}$ free to vary and found acceptable fits with temperatures $T\pdx{cool}<63-84$ eV (the range corresponds to the different thermal emission model used, i.e. \textsc{nsmaxg} or a blackbody), yielding bolometric luminosities below $L\pdx{cool}<(5-16)\times10^{32}$~\lum.
As seen in other neutron stars with ages of $\sim$1--10 kyr (e.g. PSR B0833$-$45, \citealt{2001ApJ...552L.129P}; PSR B1706$-$44, \citealt{2004ApJ...600..343M}; PSR B2334$+$61, \citealt{2006ApJ...639..377M}), this luminosity
is lower than  predicted by standard cooling curves (in the range $2\times10^{33}-10^{34}$ \lum), but can be explained with the presence of iron envelopes and/or the activation of fast cooling processes  \citep{pot20}.

The radius of the emitting region that we infer from our models, $R\pdx{em}\sim5-6$ km, is larger than expected in the framework of external re-heating, where the hot spot should have the size of the magnetic polar cap, or even smaller \citep[e.g.]{2015MNRAS.447.2631V}. We note that this discrepancy cannot be solved by different assumptions on the pulsar radius or distance.
To reconcile the observed flux with an emitting area of size comparable to the   polar cap, in fact, the star radius should be greater than 20 km \textit{and} the pulsar closer than 0.5 kpc. This seems a rather unlikely possibility, also considering the large absorption of $\gtrsim$8$\times 10^{21}$ \col required to fit the X-ray spectrum. Note that the total column density in this direction is $\sim$1.1$\times 10^{22}$ \col \citep{2016A&A...594A.116H} and the extinction maps of \citet{2019ApJ...887...93G} show that significant reddening occurs only for stars at about 1 kpc distance.
The large PF further argues against emission from the entire star surface, as it would be expected in the case of secular cooling. We note, however, that the thermal surface map of a cooling neutron star is strongly affected by the topology of the magnetic field inside the crust and can be highly inhomogeneous, especially if a strong toroidal field is present (see e.g. \citealt{2006A&A...457..937G}).

Finally, we note that evidence for large emitting regions, hotter than the remaining part of the surface, has been found also in other pulsars of age  similar to \psrj \citep{2010ApJ...725L...6C,2017A&A...597A..75M,aru18,2020MNRAS.493.1874D}. Some of them, as PSR J0007$+$7303, have also a high PF, reinforcing the hypothesis of a localized origin of this thermal component.  This seems less evident in pulsars of $\sim$10$^5$ yr, and certainly does not apply to pulsars with $\tau$ $\gtrsim$ $10^6$ yr, which have hot spots with dimensions consistent or even lower than those of the dipole polar caps \citep{rig18}. Despite the estimate of the emitting region size depends on the thermal model used, on the effects of geometrical projection, and  on the uncertainties on the distance, the possible age dependence of the thermal emission size is potentially of interest and worth of being more investigated.

\begin{acknowledgements}
This work has been partially supported through the INAF ``Main-streams'' funding grant (DP n.43/18). MR, SM and RT acknowledge financial support from the Italian Ministry for University and Research through grant 2017LJ39LM ``UNIAM''. RT acknowledges financial support from the Italian Space Agency (grant 2017-12-H.0).
\end{acknowledgements}

\bibliographystyle{aa}
\bibliography{aa39774}

\end{document}